\documentclass[compsoc,conference,a4paper,10pt,times]{IEEEtran}
\usepackage{booktabs}
\usepackage[T1]{fontenc} 
\usepackage{amsmath}
\usepackage{rotating}
\usepackage{epsfig,endnotes}
\usepackage[colorinlistoftodos,prependcaption,textsize=tiny]{todonotes}
\usepackage{url}

\usepackage{mathtools}
\usepackage[noend]{algpseudocode}
\usepackage{tabularx,colortbl}
\usepackage{multirow}
\usepackage[Symbolsmallscale]{upgreek}
\usepackage{algorithm}
\usepackage{algpseudocode}
\usepackage{mathtools}
\usepackage{tablefootnote}
\usepackage{siunitx}
\usepackage{listings}
\usepackage{enumitem}
\usepackage{color} 
\usepackage{footmisc}
\usepackage{amsmath}
\usepackage{hyperref}
\usepackage{booktabs}
\usepackage{subcaption}

\usepackage{mathtools, nccmath}
\usepackage{titlesec}

\title{bAdvertisement: Attacking Advanced Driver-Assistance Systems Using Print Advertisements}

\author{
Ben Nassi, Jacob Shams, Raz Ben Netanel, Yuval Elovici\\
Software and Information Systems Engineering, Ben-Gurion University of the Negev\\
\{nassib, jacobsh, razx, elovici\}@post.bgu.ac.il\\
\textbf{Video 1 -} \url{https://youtu.be/J4w5fpNhrs8}\\
\textbf{Video 2 -} \url{https://youtu.be/4Zj9NkTcPfs}\\
}

\begin{document}
\maketitle

\begin{abstract}
In this paper, we present bAdvertisement, a novel attack method against advanced driver-assistance systems (ADASs). bAdvertisement is performed as a supply chain attack via a compromised computer in a printing house, by embedding a "phantom" object in a print advertisement. When the compromised print advertisement is observed by an ADAS in a passing car, an undesired reaction is triggered from the ADAS. We analyze state-of-the-art object detectors and show that they do not take color or context into account in object detection. Our validation of these findings on Mobileye 630 PRO shows that this ADAS also fails to take color or context into account. Then, we show how an attacker can take advantage of these findings to execute an attack on a commercial ADAS, by embedding a phantom road sign in a print advertisement, which causes a car equipped with Mobileye 630 PRO to trigger a false notification to slow down. Finally, we discuss multiple countermeasures which can be deployed in order to mitigate the effect of our proposed attack.
\end{abstract}

\section{Introduction}

Advanced driver-assistance systems are defined as \textit{"vehicle-based intelligent safety
systems which could improve road safety in terms of crash avoidance, crash severity mitigation, and protection and post-crash phases"} \cite{ADAS-mandatory-EU}. 
ADASs are integrated into cars that range from no automation (level 0) to fully automated systems (level 5) \cite{driving2014levels}. 
ADASs consist of sensors, actuators, and decision-making algorithms which are used to initiate notifications to drivers, alerts, and steer/stop the car, depending on the automation level of the car. 

\begin{figure}[h]
\centering
\begin{subfigure}{0.5\columnwidth}
  \centering
  \includegraphics[width=0.92\columnwidth]{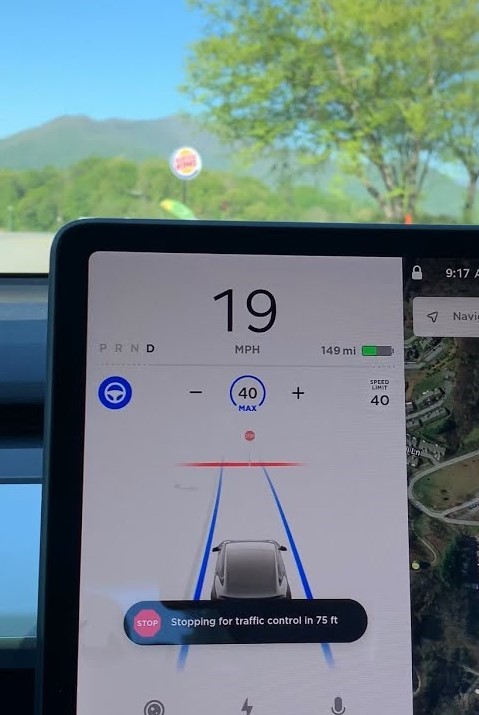}
\end{subfigure}%
\begin{subfigure}{0.5\columnwidth}
  \centering
  \includegraphics[width=0.9\columnwidth]{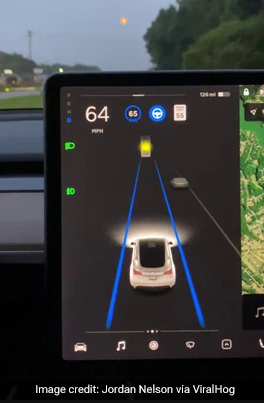}
\end{subfigure}
\caption{Left: A Tesla ADAS interpreting a Burger King sign as a stop sign. Right: A Tesla ADAS interpreting the moon as a yellow traffic light.}
\label{fig:burgerkingandmoon}
\end{figure}

Despite the ever-evolving capabilities of ADASs, in some circumstances, they still exhibit unexpected behavior. 
For example, a Tesla vehicle was shown to interpret a Burger King sign as a stop sign and slowed to a stop in response \cite{Eliot.2020}. 
In another well-known incident, a Tesla vehicle interpreted the moon as a yellow traffic light and attempted to slow down in response \cite{Levin.2021} (see Figure \ref{fig:burgerkingandmoon}).
While, these cases have proven that ADASs may unintentionally misperceive their physical surroundings, they also raise an interesting question: could an attacker intentionally cause an ADAS to misperceive the physical surroundings and trigger an undesired reaction from the ADAS? 

% A previous study \cite{phantoms} proposed to conduct such an attack through embedding a phantom in an advertisement on a digital billboard. In response to this, manufacturers have increased the security level of their digital billboards, increasing the difficulty of practically conducting this attack, which requires remotely attacking a digital billboard [TODO need a reference for increased security?].

A previous study \cite{phantoms} demonstrated such an attack by embedding a phantom road sign in an advertisement presented on a digital billboard.
The study demonstrated the split-second embedding of a road sign into a digital advertisement and showed that Tesla's autopilot triggered the car to stop in the middle of the road in response to the stop sign detected in the digital advertisement.
This study increased understanding and awareness of attacks that can be applied via digital billboards. 
The risks demonstrated in the study should serve to encourage manufacturers to increase the security level of their digital billboards, in order to prevent attackers from executing such attacks. In light of the readily available countermeasures against attacks on digital billboards, we wonder whether an attacker could find alternative ways of performing visual phantom attacks on ADASs that are not mitigated by such countermeasures.

\begin{figure}[h]
    \centering
    \includegraphics[width=0.95\columnwidth]{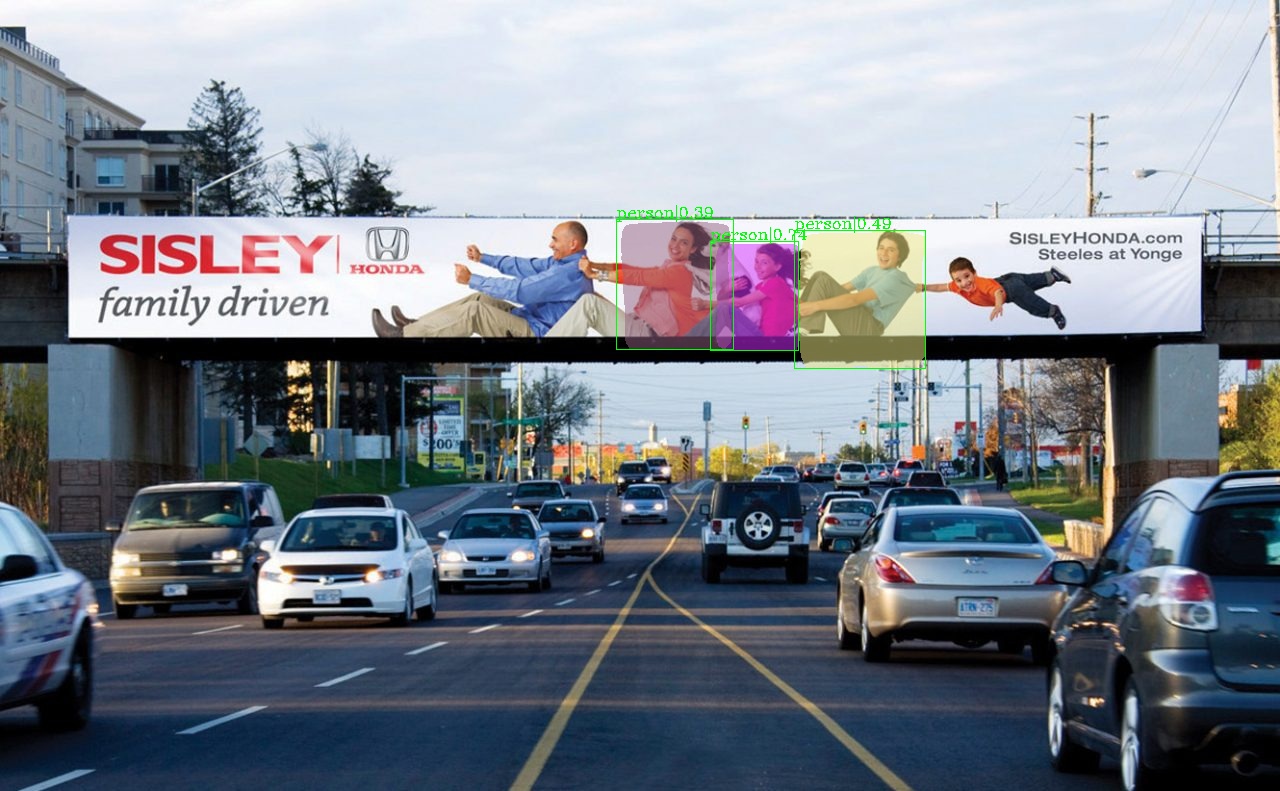}
    \caption{The state-of-the-art object detector RCNN falsely identifying people in a billboard advertisement as pedestrians.}
    \label{fig:rcnn}
\end{figure}

In this paper, we present bAdvertisement, a phantom attack on ADASs based on a phantom stealthily embedded in a print advertisement (e.g., an ad on the side/back of a bus).
Print advertisements offer significant potential for use as an attack vector. State-of-the-art object detectors were shown to falsely identify images of people in advertisements as actual pedestrians, as seen in Figure \ref{fig:rcnn}.
In the proposed attack, the attacker performs a supply chain attack, targeting the printing house where an advertisement is printed. 
The attacker manipulates the advertisement such that it is printed with an embedded phantom designed to be undetectable to humans and perceived as a traffic sign by ADASs. 
Once the advertisement is in place, an ADAS that observes it should interpret it as a traffic sign and act accordingly.

We demonstrate the use of bAdvertisement on Mobileye 630 PRO.
First, we analyze state-of-the-art object detectors and show that they do not take sign color or context into account in the process of object detection.
We then reproduce this behavior in Mobileye 630 PRO and show that its detection algorithm does not take sign color or context into account in object detection. 
Based on our findings, we show how attackers can embed a phantom traffic sign in a print advertisement by modifying it so that specific elements in the ad are interpreted as legitimate road signs. 
Finally, we provide an end-to-end demonstration of the attack: we select a random lottery advertisement, embed a phantom image of a speed limit sign in the ad, print the advertisement, and adhere it to the back of a car. We equip a second car with Mobileye 630 PRO and observe how the Mobileye camera interprets the advertisement as a legitimate traffic sign.

In this paper, we make the following contributions:
\begin{enumerate}
    % \item We propose a novel phantom attack against ADAS with an alternate attack vector, compared to prior work.
    % \item We raise awareness regarding a new target for potential attackers: advertisement printing houses.
    \item We present a novel, remote phantom attack against ADAS sensors, by embedding a phantom in an advertisement at the printing house, a new attack vector with regard to prior work. The attack raises awareness of a new target for potential attackers: printing houses that produce advertisements.
    \item The attack presented in this study is stealthy and undetectable to human drivers. In addition, the presence of the attacker is not required at the ADAS attack site.
\end{enumerate}

The structure of the paper is as follows: We review related work in Section \ref{section:related_work} and present the threat model in Section \ref{section:threat_model}.
In Section \ref{section:analysis}, we describe the experiments that showed that Mobileye 630 PRO does not take an object's color and context into account during the object detection process, and in Section \ref{section:evaluation}, we demonstrate how attackers can apply bAdvertisement against Mobileye 630 PRO. 
In Section \ref{section:countermeasures}, we discuss potential countermeasures, and we share our conclusions and plans for future work in Section \ref{section:conclusions}.

\section{Background \& Related Work}
\label{section:related_work}

Early computer vision studies aimed at developing computerized driver intelligence appeared in the mid-80s, when scientists first demonstrated a road following robot \cite{wallace1985first}. 
Studies performed from the mid-80s until 2000 established the fundamentals for the development of automated driver intelligence in related tasks, including pedestrian detection \cite{zhao2000stereo}, lane detection \cite{bertozzi1998gold}, and road sign detection \cite{de1997road}. 
However, in these early years, the vast majority of computer vision algorithms aimed at detecting objects required the developers to manually program the dedicated features in order to achieve their goal.

The increased computational power available in recent years changed the way AI models are created: an object's features are automatically extracted by training various neural network architectures on raw data.
The new automatic feature extraction approach outperformed and replaced the traditional approach that required developers to manually program an object's features. 
Motivated by AI's improved performance, the automobile industry deployed the object detectors used in today's commercial ADASs, where they are used to warn drivers of imminent collisions with pedestrians or cars, recognize road signs, etc. 
Object detectors' integration into commercial ADASs encouraged scientists to investigate the difference between a human's perception and the perception of ADASs. 

\begin{figure*}
    \centering
    \includegraphics[width=0.8\textwidth]{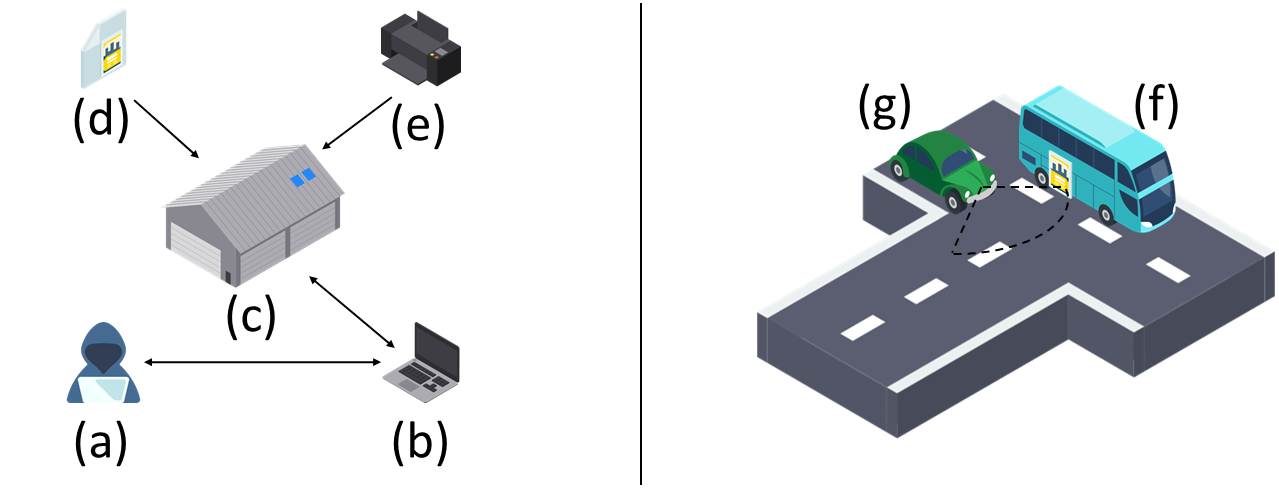}
    \caption{Threat Model: An attacker (a) infiltrates a computer (b) in a printing house's organizational network (c). The attacker embeds a phantom street sign into an advertisement file (d) before printing the ad (e). After printing and placing the advertisement in its designated location (f), an ADAS in a passing automobile (g) interprets the phantom sign as a legitimate sign and responds accordingly.}
    \label{fig:threatmodel}
\end{figure*}

In this context, visual spoofing attacks were demonstrated by \cite{sitawarin2018darts, eykholt2017robust, song2018physical,chen2018shapeshifter, Zhao-CCS2019,KeenLabs, morgulis2019fooling, phantoms}. 
Several studies have demonstrated adversarial machine learning attacks in the physical world against traffic sign recognition algorithms \cite{sitawarin2018darts, eykholt2017robust, song2018physical,chen2018shapeshifter, Zhao-CCS2019} by: (1) embedding two traffic signs in one with a dedicated array of lenses that causes a different traffic sign to appear depending on the angle of view \cite{sitawarin2018darts}, and (2) adding a physical artifact (e.g., stickers, graffiti) that looks innocent to the human eye but can mislead traffic sign recognition algorithms \cite{eykholt2017robust, Zhao-CCS2019, chen2018shapeshifter, song2018physical}. 
Other studies demonstrated adversarial machine learning attacks in the physical world against commercial ADASs by: (1) printing traffic signs that contain negligible changes that fooled the road sign detection mechanism of HARMAN's ADAS \cite{morgulis2019fooling}, (2) placing stickers on the road that caused (a) Telsa’s autopilot to deviate into the lane of oncoming traffic \cite{KeenLabs} and (b) Comma.ai's ADAS to detect dirt as a real lane \cite{274691}, (3) embedding a stop sign for a split second in a McDonald's digital advertisement presented on a digital billboard which caused Telsa’s autopilot to suddenly stop the car in the middle of the road \cite{phantoms}.
Our proposed attack, bAdvertisement, is another example of an attack against a commercial ADAS (Mobileye 630 PRO). 

Other attacks against car sensors were demonstrated by \cite{yan2016can, GPS-Spoofing-Regulus, petit2015remote, cao2019adversarial, LIDAR-USENIX20}. 
One study \cite{yan2016can} demonstrated spoofing and jamming attacks against Tesla's radar and ultrasonic sensors which caused the car to misperceive the distance to nearby obstacles. Another study \cite{GPS-Spoofing-Regulus} showed that GPS spoofing can cause Tesla’s autopilot to navigate in the wrong direction. 
Black-box and white-box adversarial attacks on LiDAR were recently presented by \cite{petit2015remote, cao2019adversarial, LIDAR-USENIX20}.

\section{Threat Model}
\label{section:threat_model}
In this section, we define phantoms, describe the attacker, present bAdvertisement, and discuss its significance. 

\textbf{Phantom:} A phantom is a two-dimensional representation of an object, used to deceive ADASs and trigger an undesired reaction. The reaction of an ADAS varies depending on its level of sophistication: a level 0-1 ADAS could issue a false notification or alarm, and a level 2 or higher ADAS could react by taking dangerous actions to avoid a perceived threat, such as sudden braking or swerving. 
Such undesired reactions could result in injury to the driver or other drivers/pedestrians and/or damage to the vehicle and surrounding objects.

\textbf{Attacker:} We consider the attacker as any malicious actor who:
\begin{enumerate}
    \item intends to disrupt the safe, orderly functioning of an ADAS via a targeted attack against the computer vision and object detection mechanisms of the ADAS, and
    % \item As a result of such disruptions, causes physical damage to the driver and other drivers/passengers/pedestrians, and/or material damage to the vehicle, ADAS, and/or other surrounding vehicles/objects.
    \item as a result of the disruption caused, triggers undesired behavior in the ADAS in response to nonexistent elements which it perceives in the surrounding environment.
\end{enumerate}
Table \ref{tab:map}  maps the desired goal (e.g., traffic collision), triggered reaction of the ADAS (e.g., sudden braking), and the embedded phantom required (e.g., stop sign).

\textbf{bAdvertisement:} bAdvertisement is a  supply chain attack on a print advertisement performed prior to its printing by the printing house. 
The attacker compromises a computer connected to the printing house's organizational network (through phishing, a malvertisement, social engineering, etc.). After infiltrating the network, the attacker embeds a stealthy phantom road sign into the advertisement prior to printing, such that an ADAS observing the advertisement (e.g., on a bus, billboard, etc.) will interpret the embedded phantom as a road sign. 
Due to the phantom road sign's stealthiness, the printing house staff and drivers on the road are unaware that the advertisement contains an embedded phantom sign.
bAdvertisement exploits the fact that commercial ADASs fail to take sign color or context (surroundings) into account when identifying and classifying signs. In bAdvertisement, the phantoms used resemble an object an ADAS would consider a traffic sign, but which human drivers wouldn't notice.
After printing the manipulated advertisement with the embedded phantom sign and placing it in its designated location, any ADAS that observes the advertisement will perceive the phantom sign embedded in the advertisement and respond accordingly.

Figure \ref{fig:threatmodel} provides a visual representation of bAdvertisement's threat model, including the pre-printing stage in the printing house (a-e) and the post-printing stage on the road (f-g). 
We note that the attacker's involvement is only required in the pre-printing stage.

\begin{table}
\centering
\caption{Mapping an attack to a desired result.}
\label{tab:map}
\resizebox{1.0\columnwidth}{!}{%
\begin{tabular}{|l|l|l|}
    \hline
    \textbf{Desired Result} & \textbf{\begin{tabular}[c]{@{}l@{}}Triggered Reaction\\from the ADAS\end{tabular}} & \textbf{\begin{tabular}[c]{@{}l@{}}Type of \\ Phantom\end{tabular}} \\
    \hline
    
    \multirow{3}{*}{\begin{tabular}[c]{@{}l@{}}Traffic \\ collision\end{tabular}} & \multirow{3}{*}{\begin{tabular}[c]{@{}l@{}}Sudden braking\end{tabular}} & Stop sign \\
    \cline{3-3}
     &  & No entry sign  \\ \cline{3-3} 
     &  & \begin{tabular}[c]{@{}l@{}}Obstacle (e.g., car)\end{tabular} \\ \hline
    \begin{tabular}[c]{@{}l@{}}Reckless/illegal \\ driving behavior\end{tabular} & \begin{tabular}[c]{@{}l@{}}Fast driving\end{tabular} & Speed limit \\
    \cline{1-3}
    \multirow{2}{*}{\begin{tabular}[c]{@{}l@{}}Traffic jam\end{tabular}} & \begin{tabular}[c]{@{}l@{}}Decreasing driving speed \end{tabular} & Speed limit  \\ \cline{2-3}
     & Stopping & No entry sign  \\ \cline{1-3}
    \begin{tabular}[c]{@{}l@{}}Directing traffic\\ to specific roads\end{tabular} & \begin{tabular}[c]{@{}l@{}}Avoiding certain roads\end{tabular} & No entry sign  \\ \hline
\end{tabular}
}
%\vspace{-1.7em}
\end{table}

\textbf{Significance:} The significance of bAdvertisement compared to classical adversarial attacks \cite{sitawarin2018darts, eykholt2017robust, song2018physical, chen2018shapeshifter, Zhao-CCS2019, morgulis2019fooling} and phantom attacks \cite{phantoms} are:
\begin{enumerate}
    \item bAdvertisement can be applied remotely by embedding the phantom in the advertisement before the printing process, and it does not require the attacker's presence at the attack site, unlike previous attacks.
    \item The targeted computers connected in the organizational network of the printing house are an easier target to compromise than a digital billboard, due to increased awareness of the risks posed by unsecured digital billboards demonstrated in a prior study \cite{phantoms}.
\end{enumerate}

\section{Analysis}
\label{section:analysis}
\begin{figure}[h]
\centering
\includegraphics[width=1.0\columnwidth]{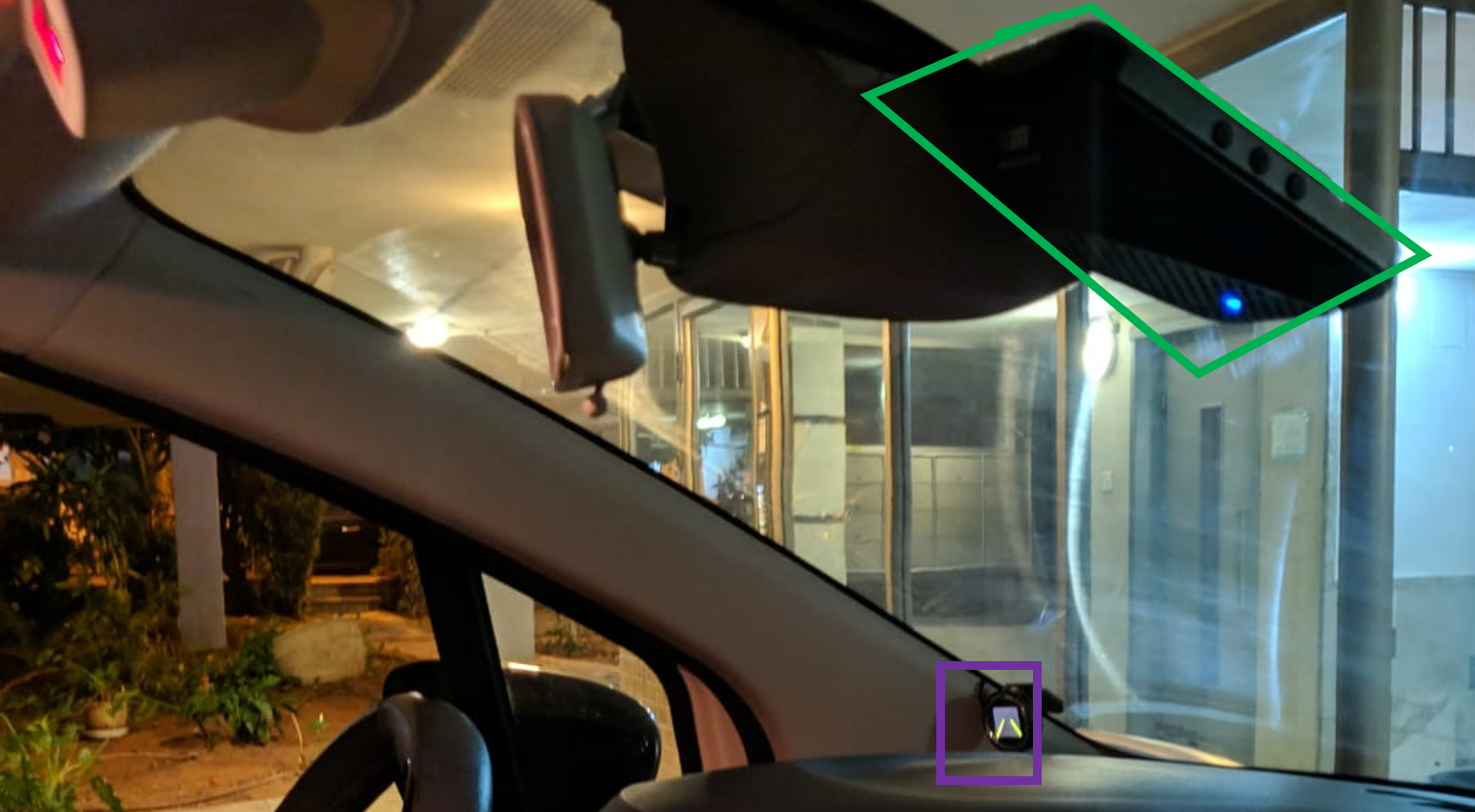}
\caption{Mobileye 630 PRO components: video camera (framed in green) and UI screen (framed in purple).}
\label{fig:mobileye}
\end{figure}

In this section, we analyze the sensitivity of state-of-the-art object detectors to color changes in projected road signs, as well as their sensitivity to sign context, and reproduce these findings with Mobileye 630 PRO. 
Mobileye 630 PRO is an external ADAS that provides function-specific vehicle automation (Level 0). Mobileye 630 PRO contains two main components, as can be seen in Figure \ref{fig:mobileye}. 
The first component is a camera, which is installed on the windshield, under the rear view mirror, and the second component is a small display that is placed in front of the driver and provides visual and auditory alerts about the surroundings as needed. 
Mobileye 630 PRO has the following features: (1) Lane departure warning. This feature is activated when a lane deviation occurs without proper signal notification by the driver; this feature is activated when the driving speed is over 55 kilometers per hour. 
(2) Pedestrian collision warning. This feature notifies the driver of an imminent collision with a pedestrian or cyclist; this feature is only used in daylight and is activated when the driving speed is under 50 km/h (this feature can be configured to a value of up to 70 km/h).
(3) Forward collision warning. This feature notifies the driver about rear-end collisions with any type of vehicle. 
(4) Headway monitoring and warning. This feature notifies the driver when there is an unsafe distance between the driver's vehicle and the vehicle ahead of it; this feature is activated when the driving speed is over 30 km/h. (5) Upcoming traffic sign recognition. This feature recognizes and reads traffic signs (speed limit, entering a highway, etc.) and notifies the driver of upcoming traffic signs via its visual display.
The next two subsections address Mobileye 630 PRO's traffic sign recognition abilities and more specifically, its reliance on color and context for traffic sign recognition.

\subsection{Color Sensitivity}

\begin{figure}[h]
\centering
\includegraphics[width=0.6\columnwidth]{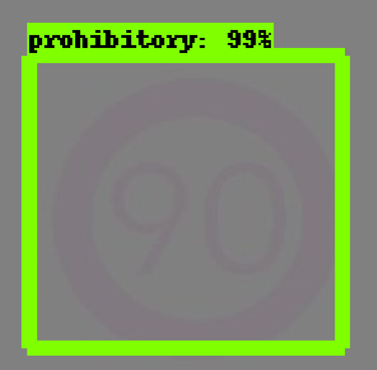}
\caption{The Faster\_rcnn\_inception\_v2 \cite{ARCOSGARCIA2018332} object detector classifies a phantom traffic sign, created by lowering the green component of the gray background, as legitimate.}
\label{fig:graysign}
\end{figure}

Many state-of-the-art object detectors are not sensitive to color and do not consider it when identifying and classifying objects.  
As can be seen in Figure \ref{fig:graysign}, the object detector Faster\_rcnn\_inception\_v2 \cite{ARCOSGARCIA2018332} recognizes a gray phantom, created by reducing the value of the green component of a gray background, as a legitimate traffic sign. This experiment is aimed at assessing Mobileye 630 PRO's color sensitivity with regard to its traffic sign recognition capabilities.

\begin{figure}[h]
    \centering
    \includegraphics[width=0.95\columnwidth]{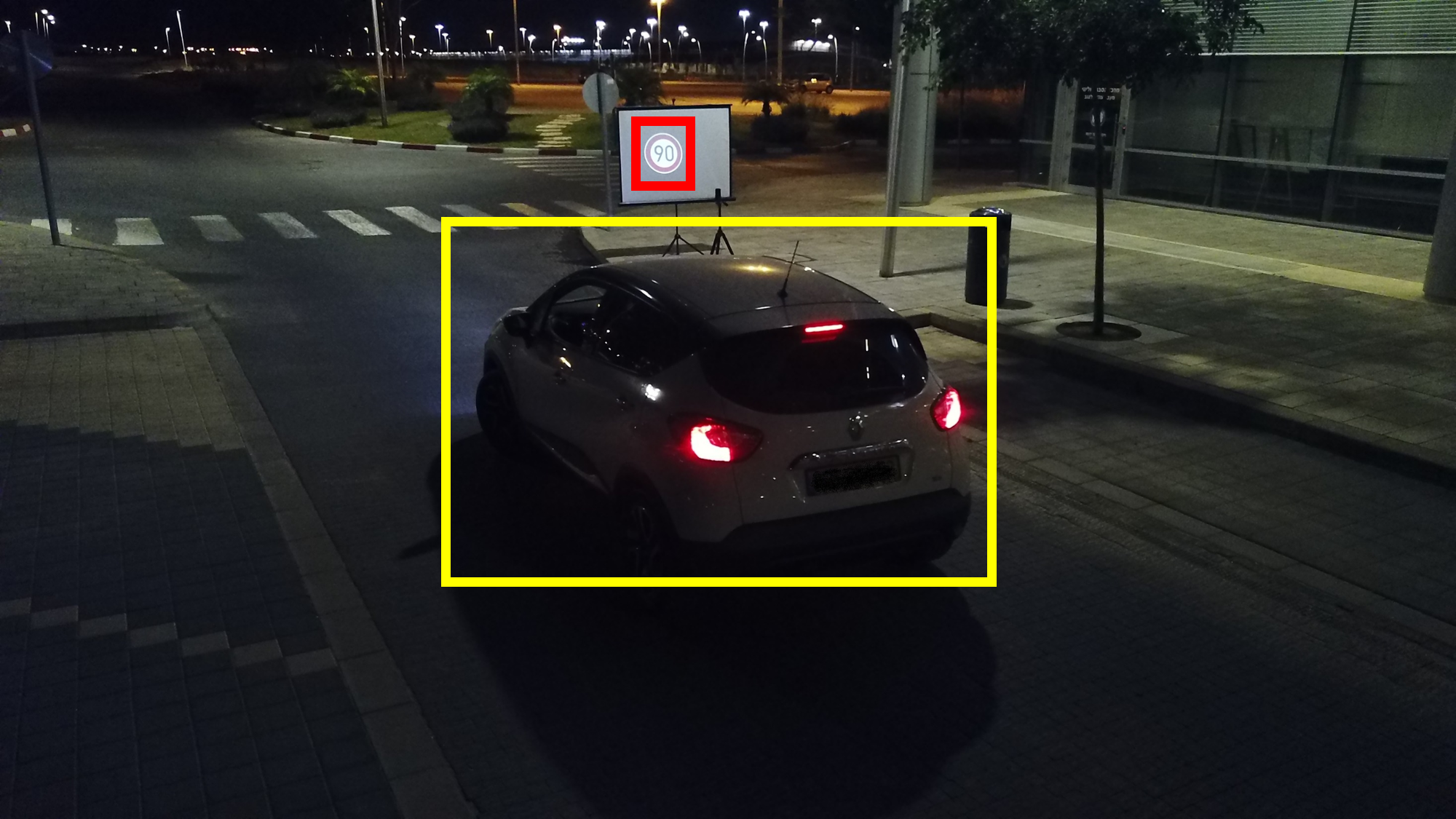}
    \caption{The experimental setup. The projected sign, and attacked car are respectively framed in red and yellow.}
    \label{fig:exp_set}
\end{figure}

\begin{figure}[h]
\centering
\includegraphics[width=0.95\columnwidth]{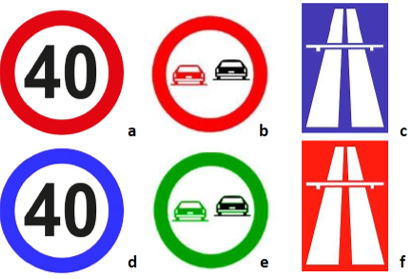}
\caption{Examples of different colored traffic signs. Signs a-c are displayed with their true colors, while signs d-f are displayed with modified colors.}
\label{fig:coloredsigns}
\end{figure}

\textit{1) Experimental Setup:} In this analysis, a white projector screen served as the surface for projecting the traffic signs. 
The signs were projected by a portable projector situated on a tripod which was located 2.5 meters away from the screen; the projections were directed toward the center of the screen. 
While the projections were directed toward the screen, we drove a car (Renault Captur), equipped with Mobileye 630 PRO, in a straight line, toward the screen, at an approximate speed of 25 km/h on a closed road. 
The experimental setup can be seen in Figure \ref{fig:exp_set}.
We initially verified that the ADAS detected the signs when they were presented with their true colors. We then projected the signs using different colors.
Examples of different sign colors, as seen in Figure \ref{fig:coloredsigns}, include the signs' respective true colors (a-c), as well as modified colors which aren't used in actual traffic signs (d-f). 

\textit{2) Results:} Mobileye 630 PRO's response to the projections demonstrate that it is not sensitive to the color of the sign, since all of the projected signs were interpreted as real signs, regardless of their color.

\textit{3) Conclusion:} Based on these results, we can conclude that the Mobileye 630 PRO camera classifies a sign according to its shape and does not take color into account.

\subsection{Context Sensitivity}

\begin{figure}[h]
    \centering
    \includegraphics[width=0.95\columnwidth]{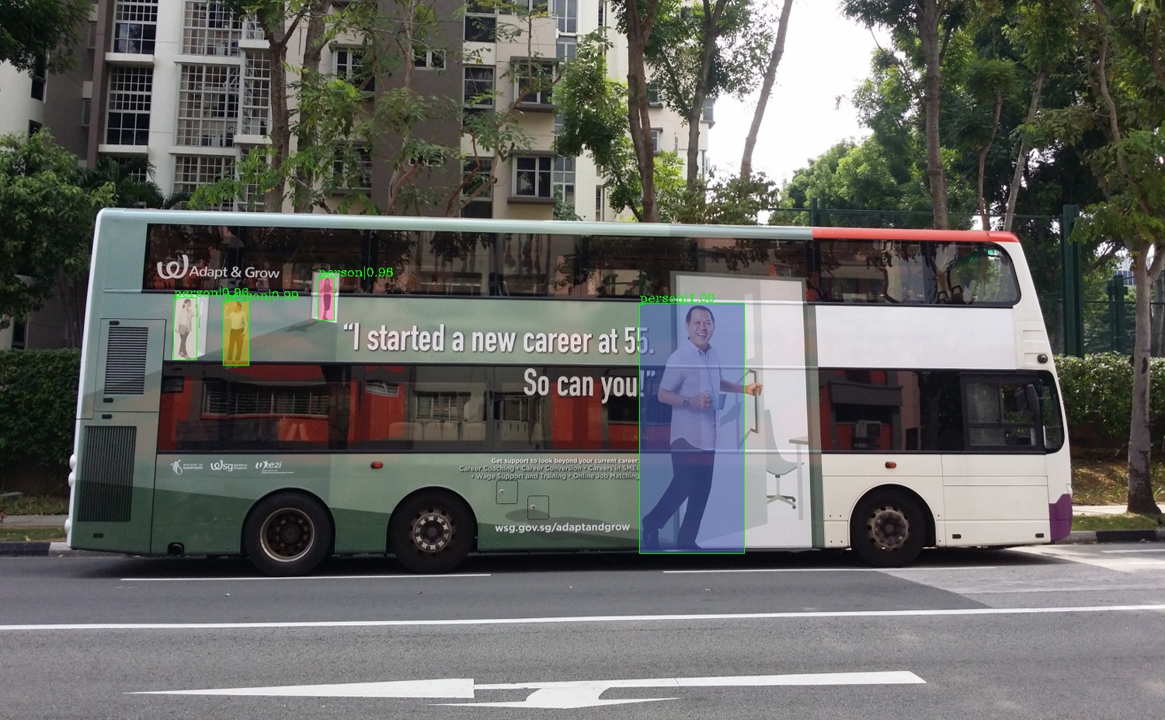}
    \caption{The ResNet object detector falsely identifying people in a bus advertisement as pedestrians.}
    \label{fig:resnet}
\end{figure}

Similarly, many state-of-the-art object detectors fail to take context into account when identifying and classifying objects. 
As can be seen in Figure \ref{fig:resnet}, the ResNet object detector recognized an image of a person on a bus advertisement as an actual person.

\textit{1) Experimental Setup:} In order to gauge the reliance of Mobileye 630 PRO's camera on context, we projected traffic signs onto a tree, so that the signs were projected on the leaves of the tree. The projector was situated on a tripod located 2.5 meters away from the tree. While the projections were directed at the leaves, we drove a car (Renault Captur), equipped with Mobileye 630 PRO, in a straight line, toward the tree, at an approximate speed of 25 km/h on a closed road.

\begin{figure}[h]
\centering
\includegraphics[width=0.6\columnwidth]{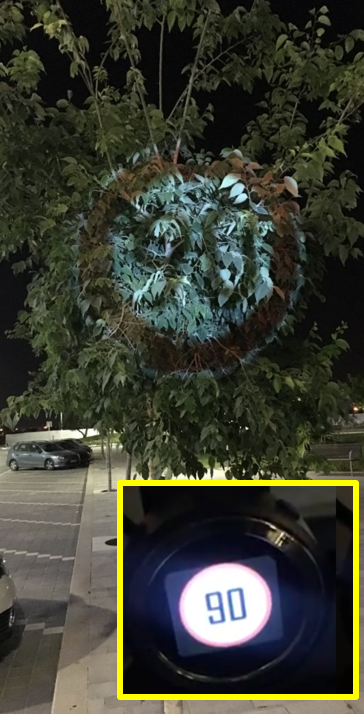}
\caption{The Mobileye 630 PRO camera interprets the projection of a sign on a tree as a legitimate traffic sign.}
\label{fig:treeprojection}
\end{figure}

\textit{2) Results:} Mobileye 630 PRO's camera falsely identified the projections on the leaves as legitimate traffic signs, as can be seen in Figure \ref{fig:treeprojection}, indicating that it is not sensitive to the context of the sign.

\textit{3) Conclusions:} Based on these results, we can conclude that Mobileye 630 PRO classifies a sign according to its shape and does not take a sign's context into account.

Our findings show that Mobileye 630 PRO does not consider color or context, a fact which can be exploited by attackers in order to execute bAdvertisement.

\section{Evaluation}
\label{section:evaluation}
In this section, we demonstrate an attacker's ability to cause Mobileye 630 PRO's camera to incorrectly classify phantoms embedded in a print advertisement as legitimate road signs.

\subsection{Experimental Setup}

\begin{figure}[h]
    \centering
    \includegraphics[width=0.95\columnwidth]{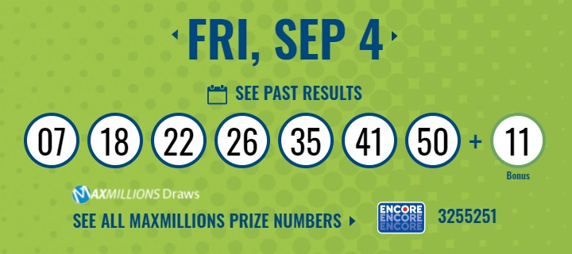}
    
    \vspace{1.5em}
        \includegraphics[width=0.95\columnwidth]{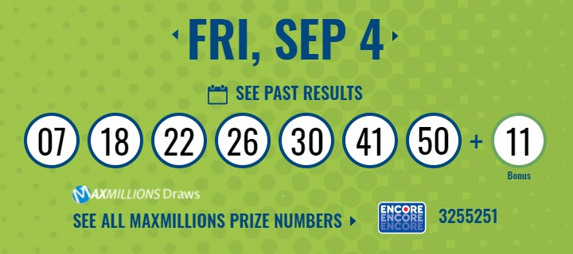}
\caption{Top: The original lottery advertisement. Bottom: The lottery advertisement after embedding the phantom street sign. As can be seen, the number 35 in the original advertisement was replaced with the number 30 in the modified advertisement.}
\label{fig:advertisements}
\end{figure}

\begin{figure}[h]
\centering
\includegraphics[width=0.95\columnwidth]{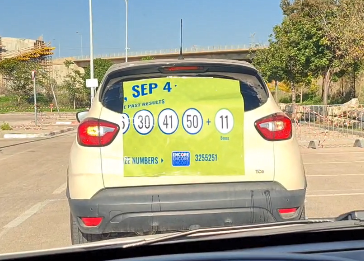}
\caption{The modified print advertisement attached to a car, to mimic an advertisement on the back of a bus.}
\label{fig:adoncar}
\end{figure}

We started by selecting an advertisement in which to embed a phantom traffic sign. 
The advertisement chosen was a lottery advertisement presenting winning numbers, since the winning numbers bear a strong similarity to a speed limit sign. We then added a phantom speed limit sign in the advertisement by changing the middle number from 35 to 30, since Mobileye 630 PRO doesn't recognise 35 km/h speed limit signs, but does recognise 30 km/h speed limit signs. It should be noted that, in addition to the embedded 30 km/h phantom speed limit sign, the advertisement already contains a number which could be recognised as a speed limit sign by Mobileye 630 PRO (50). The modifications made were stealthy and unobservable to humans, without any indication that the ad was manipulated or that a phantom street sign was embedded.
The original and modified advertisements are presented in Figure \ref{fig:advertisements}. 
The advertisement was printed and attached to a car, in order to mimic an advertisement on the back of a bus, as can be seen in Figure \ref{fig:adoncar}. 
In this experimental setup, the target vehicle (a Renault Captur with Mobileye 630 PRO installed) followed the car with the printed advertisement in a closed parking lot. 
The advertisement was positioned so that it could be detected by the Mobileye 630 PRO camera. 
We demonstrate two scenarios in this experimental setup. One scenario is an attempt to slow down a car with an ad containing a 30 km/h phantom traffic sign, and the second scenario is an attempt to accelerate a car with an ad containing a 50 km/h phantom traffic sign.
Videos of the demonstrations were uploaded to an anonymous YouTube account and can be seen via the video links provided at the beginning of the paper.

\subsection{Results}
\begin{figure}[h]
\centering
\begin{subfigure}{\columnwidth}
    \centering
    \includegraphics[width=0.95\columnwidth]{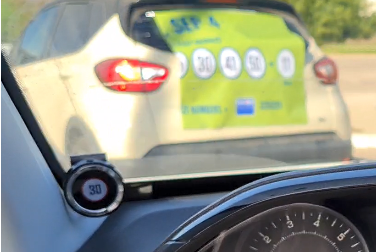}
    \caption{}
    \label{fig:30signid}
\end{subfigure}
\begin{subfigure}{\columnwidth}
    \centering
    \includegraphics[width=0.95\columnwidth]{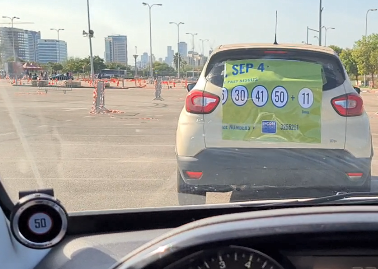}
    \caption{}
    \label{fig:50signid}
\end{subfigure}
\caption{Top: Mobileye 630 PRO sensor identifying a 30 km/h sign in the lottery advertisement. Bottom: Mobileye 630 PRO sensor identifying a 50 km/h sign in the lottery advertisement.}
\label{fig:signid}
\end{figure}

\textit{First Scenario (30 km/h sign):}
When the target vehicle followed the car with the advertisement attached to it, the Mobileye 630 PRO camera interpreted the advertisement as a 30 km/h sign (see Figure \ref{fig:30signid}). 
This interpretation can be attributed to the Mobileye camera's perception of the modified lottery number as a legitimate 30 km/h sign.

\textit{Second Scenario (50 km/h sign):}
While following the car with the displayed advertisement, the Mobileye 630 PRO camera interpreted the advertisement as a 50 km/h sign (see Figure \ref{fig:50signid}), attributed to interpreting the lottery number second from the right as a 50 km/h speed limit sign. 

These results show that the bAdvertisement attack is capable of misleading ADASs with phantom traffic signs in print ads, causing cars to: (1) slow down below the actual speed limit, as well as (2) speed up beyond the actual speed limit. 

\subsection{Conclusions}
These results corroborate our hypothesis that targeted embedding of phantoms in print advertisements presents an effective attack vector which can be used to trigger undesired reactions from an ADAS, like false notifications to the driver about their surroundings or sudden dangerous swerving or stopping in response to a nonexistent threat. 
Such responses could prove dangerous to the driver and passersby, as well as to surrounding cars and property.
The experiment demonstrates how an attacker can cause traffic congestion by slowing down the speed of a group of cars by embedding a phantom road sign in an advertisement placed on a bus. Embedding a speed limit sign lower than the actual speed limit would cause all ADASs in the vicinity of the advertisement to reduce their cars' speed to match the phantom sign. Traffic would therefore slow down, and the traffic would become congested.

\section{Countermeasures}
\label{section:countermeasures}
In this section, we discuss potential countermeasures that could be used to mitigate the effect of the bAdvertisement attack. 
These countermeasures help ensure that ADASs can correctly identify and verify the legitimacy of objects they perceive as traffic signs and obstacles.

\begin{figure*}[h]
\centering
\includegraphics[width=0.8\linewidth]{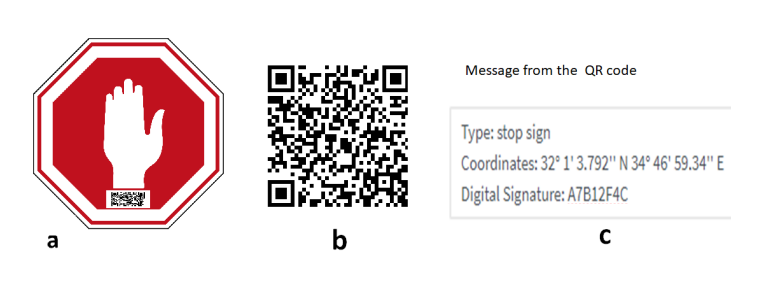}
\caption{A potential countermeasure against our proposed attack. (a) A traffic sign embedded with a QR code; (b) A close-up of the embedded QR code; (c) The message extracted by the ADAS from the QR code, including content, location, and information for authentication.}
\label{fig:qrcodecounter}
\end{figure*}

One potential countermeasure against our proposed attack method is to embed a QR code on legitimate traffic signs. 
The QR code would be readable by the ADAS and provide verification information, enabling the ADAS to confirm a specific traffic sign's location and contents, and respond accordingly. 
Traffic signs without such a QR code would be ignored. 
However, this method is vulnerable to GPS spoofing, which could cause the ADAS to interpret the sign as a legitimate sign located elsewhere. 
Details regarding this potential countermeasure are provided in Figure \ref{fig:qrcodecounter}.

Another potential countermeasure is for ADASs to utilize sensors that detect three-dimensional objects, like LiDAR, rather than cameras, which are limited to two-dimensional perception. 
The bAdvertisement attack takes advantage of the fact that such cameras cannot differentiate between two-dimensional and three-dimensional objects and thus may perceive a two-dimensional phantom image of a traffic sign as a legitimate three-dimensional traffic sign. 
However, this countermeasure would be vulnerable to an attacker who physically deploys counterfeit three-dimensional traffic signs, which an ADAS equipped with a three-dimensional sensor would still falsely identify as a legitimate traffic sign.

Another countermeasure is to utilize a distributed database of traffic sign locations and content, where an ADAS receives updates from other drivers regarding local traffic signs. 
Such a countermeasure would bolster the ability of an ADAS to verify the legitimacy of a traffic sign, since it would receive updates from other drivers regarding the location and content of local traffic signs. 
This method is vulnerable to adversarial attacks and GPS spoofing, since it requires a sustained Internet connection and depends on input from other drivers.

In the future, when transportation infrastructure (like roads and traffic signs) is designed with autonomous vehicles in mind, ADASs may not need to rely on computer vision and object detection to gain information about the surrounding road features. 
Such information could be transmitted via a wireless communication protocol between traffic signs and an ADAS, while object detection could be relegated to detecting the sudden appearance of obstacles on the road, such as pedestrians or other vehicles. 
In such a case, attacks like bAdvertisement would lose a lot of their potential utility. 
While traffic and road information still needs to be acquired via computer vision, such countermeasures can provide an additional layer of protection against phantom attacks like bAdvertisement.

\section{Conclusions \& Future Work}
\label{section:conclusions}
In this paper, we presented bAdvertisement, a novel attack on ADASs which embeds a phantom traffic sign in a print advertisement. 
By conducting a supply chain attack on the advertisement, the ad is manipulated prior to printing, via a compromised computer at the printing house, and printed and deployed with the embedded phantom. 
We show that this attack can be used to take advantage of Mobileye 630 PRO's lack of color and context sensitivity by by modifying a lottery advertisement so that it includes an embedded phantom speed limit sign, which is positively identified by the ADAS.
In other scenarios, as shown in Figure \ref{fig:resnet}, whether the ADAS falsely identifies a person in the ad or correctly identifies the bus displaying the ad makes little difference, since the response would be the same in each case: to stop or slow the car to avoid hitting the obstacle. We present a scenario where the ADAS is tricked into responding differently than it would if an embedded phantom was not embedded in the advertisement (e.g., speeding up instead of stopping).
We also present multiple countermeasures that could serve as a defense against our proposed attack.

Future research could include identifying additional advertisement templates and phantom traffic signs that can induce undesired behavior from ADASs and introducing/implementing countermeasures in ADASs to mitigate the effectiveness of such attacks.

\bibliographystyle{IEEEtran}
\bibliography{IEEEabrv,main}

\end{document}